\documentclass[12pt]{article}

\usepackage{amsmath}
\newcommand{\ft}[2]{{\textstyle\frac{#1}{#2}}}
\newcommand{\eqn}[1]{(\ref{#1})}
\newcommand{\be}{\begin{equation}}
\newcommand{\ee}{\end{equation}}
\newcommand{\uu}{\Upsilon}
\newcommand{\ep}{\epsilon}
\newcommand{\de}{\delta}
\newcommand{\co}{{\cal O}}
\newcommand{\s}{\sigma}
\newcommand{\bu}{{\bar\Upsilon}}
\newcommand{\hc}{{\hat\chi}}

\newcommand{\bea}{\begin{eqnarray}}
\newcommand{\eea}{\end{eqnarray}}
\newcommand{\nn}{\nonumber}

\newcommand{\I}{\mathrm{i}}
\newcommand{\e}{\mathrm{e}}

\newcommand{\si}{\sigma}
\newcommand{\p}{\partial}
\newcommand{\half}{\frac{1}{2}}

\newcommand{\ie}{{\it i.e.}}
\newcommand{\eg}{{\it e.g.}}

\DeclareSymbolFont{AMSa}{U}{msa}{m}{n}
\DeclareSymbolFont{AMSb}{U}{msb}{m}{n}
\DeclareMathSymbol{\fieldR}{\mathalpha}{AMSb}{"52}

\newif\ifpdf
\ifx\pdfoutput\uundefined
\pdffalse 
\else
\pdfoutput=1 
\pdftrue
\fi

\begin{document}
\begin{titlepage}
\begin{center}
\hfill ITP-UU-04/02  \\
\hfill SPIN-04/02  \\
\hfill YITP-04/04  \\
\hfill MCTP-04/02  \\
\hfill {\tt hep-th/0402132}\\
\vskip 15mm

{\Large {\bf Quantum Corrections to the Universal Hypermultiplet
and Superspace \\[3mm] }}

\vskip 10mm

{\bf Lilia Anguelova$^{a}$, Martin Ro\v{c}ek$^{b}$ and Stefan Vandoren$^{c}$ }

\vskip 4mm
$^a${\em Michigan Center for Theoretical Physics}\\ 
{\em Randall Laboratory of Physics,
University of Michigan}\\ {\em Ann Arbor, MI 48109-1120, USA} \\
{\tt anguelov@umich.edu}\\

$^b${\em C.N. Yang Institute for Theoretical Physics}\\
{\em SUNY, Stony Brook, NY 11794-3840, USA}\\
{\tt rocek@insti.physics.sunysb.edu}\\[2mm]

$^c${\em Institute for Theoretical Physics} and {\em Spinoza
Institute}\\
{\em Utrecht University, Utrecht, The Netherlands}\\
{\tt vandoren@phys.uu.nl}\\

\vskip 6mm

\end{center}

\vskip .1in

\begin{center} {\bf ABSTRACT }\end{center}
\begin{quotation}\noindent

We investigate quantum corrections to the effective action of the universal
hypermultiplet in the language of projective superspace. We rederive the 
recently found one-loop correction to the universal hypermultiplet moduli 
space geometry. The deformed metric is described as a superspace action 
in terms of a single function, homogeneous of first degree. Our framework 
leads us to a natural proposal for the nonperturbative moduli space metric 
induced by five-brane instantons.

\end{quotation}
\vfill
\flushleft{\today}

\end{titlepage}

\eject

\tableofcontents

\section{Introduction}
\setcounter{equation}{0}

Quantum corrections to the moduli spaces of low energy effective actions of 
string compactifications are important for many reasons, chief among them
being their implications for supersymmetry breaking and the lifting of 
vacuum degeneracy. However, studying them is in general quite a difficult 
task. Explicit computations are manageable in the case of type II 
compactification on a Calabi-Yau three-fold ($CY(3)$) due to the stringent 
restrictions imposed by $N=2$, $D=4$ local supersymmetry. As the geometry
of the moduli space of hypermultiplets coupled to supergravity is the same in 
four and five dimensions, this is more than just a toy model. The five-dimensional
case is relevant for heterotic M-theory \cite{LOSW,HW}, which may have
interesting implications for string phenomenology \cite{EW} and cosmology 
\cite{LOW}.

The field content of the four-dimensional low energy limit of type II string theory is 
determined by the Hodge numbers of the internal Calabi-Yau space; in particular 
for IIA there are $h_{1,2}+1$ hypermultiplets and $h_{1,1}$ vector multiplets. The 
``$+1$'' is the universal hypermultiplet, which appears in  compactification on {\it 
any} $CY(3)$ \cite{CFG}. It contains as bosonic fields the dilaton, an axion (the 
dual of the NS-NS two-form) and a complex scalar originating from the RR 
three-form potential. As the dilaton belongs to a hypermultiplet, the vector 
multiplet moduli space is protected from quantum corrections (in the string
coupling constant), but the hypermultiplet moduli space is not. Loop corrections 
to the latter were studied in \cite{AFMN}. More precisely, these authors
investigated the non-universal directions (\ie, directions orthogonal to the 
universal hypermultiplet) and found a one-loop contribution proportional to the
Euler number of the Calabi-Yau manifold. Perturbative corrections to the universal
hypermultiplet target space itself were first considered in \cite{S} (see also
\cite{GHL}). As was stressed there, this is particularly interesting since any
corrections to this moduli space are gravitational in nature.\footnote{The reason is
that the {\bf c}-map (or T-duality) maps the universal hypermultiplet into the gravity
multiplet instead of into a vector multiplet \cite{CFG,FS}.}

Recently this issue was reconsidered in \cite{amtv} and it was
found, in contrast to the claim of \cite{S}, that there {\em is}
a one-loop correction which cannot be absorbed in a field
redefinition. In the present paper, we derive the form of this new term in the 
framework of projective superspace \cite{proj}; this leads us to propose a
natural candidate for the nonperturbative metric of the universal hypermultiplet 
moduli space induced by five-brane instantons. Recall that although since the 
work of \cite{BBS} it is known conceptually that instantons due to euclidean 
membranes and five-branes wrapping supersymmetric cycles\footnote{As 
we are considering type IIA, clearly we mean D$2$- and NS$5$-branes. 
There are no contributions from D$4$-brane instantons since $CY$ 
three-folds do not have nontrivial supersymmetric five-cycles.} 
in the $CY$ give nonperturbative corrections to the moduli space metric, 
explicit metrics have been proposed only for special cases (with vanishing 
five-brane charge) \cite{SK}. The relevance of the quantum (and in 
particular {\it nonperturbative}) corrections to the universal hypermultiplet for 
cosmological applications (more precisely, finding de Sitter vacua in $N=2$ 
gauged supergravity) was underscored recently by the work of \cite{BM}.

What enables us to use the powerful projective superspace techniques \cite{proj} 
is the recently developed approach \cite{DWKV,DWRV1,ARV} for studying 
hypermultiplet couplings to supergravity.  Instead of the usual description in 
terms of $4n$-(real) dimensional quaternion-K\"{a}hler (QK) spaces \cite{BW}, 
we use the corresponding $4(n+1)$-dimensional hyperk\"ahler cones 
(HKC) \cite{Swann,Galicki}.  As hyperk\"ahler geometry is in general simpler 
than quaternion-K\"{a}hler geometry, one finds that many calculations simplify; 
for example, in \cite{DWRV2} a simple derivation of the scalar potential of vector 
multiplets and hypermultiplets in gauged $N=2$ supergravity was given. 
Furthermore, the HKC approach has also turned out to be convenient for the 
recent construction of effective supergravity actions that incorporate the additional 
light degrees of freedom appearing in flop transitions of M-theory on $CY(3)$ 
\cite{JMS}. Most relevant for our purposes, studying hypermultiplet  couplings to 
supergravity is now translated to studying sigma models with hyperk\"ahler target 
spaces, which is exactly what projective superspace is designed for \cite{HKLR,polar}. 
A particularly interesting example, somewhat comparable to our case, is the 
projective superspace description of the Atiyah-Hitchin metric \cite{IR} and our aim 
is to discuss the universal hypermultiplet and its quantum corrections in a similar way.

The present paper is organized as follows. In Section 2 we concentrate on the 
classical geometry. To recall necessary material and establish notation, in 2.1 
we give the metric of the universal hypermultiplet target space in several 
coordinate systems along with the coordinate transformations among them, 
and identify an important set of isometries forming a Heisenberg algebra. 
In 2.2 we provide background on the HKC approach and projective superspace 
and explain the description of the classical moduli space in this framework.
In 2.3 we elaborate on the precise relationship between four-dimensional QK 
metrics with $U(1)\times U(1)$ isometries and the metrics of the corresponding 
HKCs. Section 3 is devoted to the perturbative corrections. In 3.1 we review the 
essential role of the Heisenberg algebra and the result of \cite{amtv} about the 
one-loop correction. In 3.2 we derive the projective superspace form of the 
perturbatively deformed metric by considering deformations that preserve the 
Heisenberg algebra. In 3.3 we show that the deformed HKC metric that we have 
found reproduces the result of \cite{amtv} upon reduction to the QK space. In 
Section 4 we explain our proposal for the moduli space metric 
induced by five-brane instantons. We also comment on the 
possible structure of the full nonperturbative metric. Appendix A
contains the derivation of the action of the QK Heisenberg 
algebra on the corresponding HKC. In Appendix B we give an
alternate dual description of the universal hypermultiplet in 
projective superspace.

\section{Classical aspects}
\setcounter{equation}{0}

The tree-level quaternion-K\"ahler target space for the
universal hypermultiplet is \cite{CFG,FS}
\be \label{UHM-QK}
{\cal M}_H = \frac{\mathrm{SU}(1,2)}{\mathrm{U}(2)}\ .
\ee
It is obtained from a compactification of type IIA strings on a Calabi-Yau manifold with 
$h_{1,2}=0$, \ie, without complex structure deformations. The generic low-energy 
effective action is $N=2$ Poincar\'e supergravity coupled to $h_{1,2}+1$ hypermultiplets 
as well as $h_{1,1}$ vector multiplets that play no role in the subsequent discussion.
Suppressing the Einstein-Hilbert term, the bosonic Lagrangian for the hypermultiplets 
takes the form of a nonlinear sigma-model,
\be
 e^{-1}\mathcal{L}_\mathrm{HM}=-\frac{1}{2}\,g_{AB}\,\partial_\mu \phi^A
\partial^\mu \phi^B\ ,
\ee
where for the universal hypermultiplet, $g_{AB}$ is the metric on
the coset space \eqref{UHM-QK}. Before we reformulate this action in
superspace, we first discuss some properties of the metric in 
different coordinate systems.

\subsection{Quaternionic description and isometries}

The coset manifold (\ref{UHM-QK}) is both quaternion-K\"ahler and K\"ahler,
and there exist complex coordinates $S$ and $C$ in which the K\"ahler
potential is given by
\be\label{S-C}
K=-{\rm ln} (S+\bar S -2C\bar C)\ .
\ee
The isometry group of this manifold is SU(1,2) and a particular
subgroup of isometries is generated by the Heisenberg algebra that acts as
\be \label{HeisTr}
\delta S=\I\alpha + 2\bar\epsilon\, C\ ,\qquad
\delta C=\epsilon\ .
\ee
The parameters $\alpha$ and $\epsilon$ are real and complex respectively.

An alternative parametrization of the universal hypermultiplet metric in terms of 
four real scalars is
\be\label{real-UHM}
  {\rm d}s^2=  ({\rm d}\phi)^2+
    \e^{-\phi} \big( ({\rm d} \chi)^2+ ({\rm d}\varphi)^2\big)+
\e^{-2\phi} \big( {\rm d}\si + \chi {\rm d}\varphi \big)^2 \ .
\ee
The relation with the complex coordinates above is given by
\begin{eqnarray}\label{real-var}
 \e^\phi=\half (S+\bar S - 2 C \bar{C}) & \qquad & \chi = C+\bar C\
 ,\nonumber\\
 \sigma=\frac{\I}{2}(S-\bar S + C^2 - \bar{C}^2) & \qquad & \varphi =
 -\I (C-\bar C)\ .
\end{eqnarray}
In this basis, the Heisenberg algebra acts as
\be\label{H-A}
\delta \phi =0\ ,\qquad  \delta \chi = \epsilon + \bar \epsilon\ ,\qquad
\delta \sigma = -\alpha -  (\epsilon + \bar \epsilon )\varphi\ ,\qquad
\delta \varphi = -\I (\epsilon -\bar \epsilon)\ .
\ee
The generators satisfy the commutation relations,
\be\label{Heis-alg}
[\delta_\alpha,\delta_\epsilon]=[\delta_\alpha,\delta_{\bar \epsilon}]=0
\ ,\qquad [\delta_\epsilon, \delta_{\bar \epsilon}]=
2i \delta_{\alpha(\epsilon, \bar \epsilon)}~,\qquad \alpha(\epsilon, \bar \epsilon) 
\equiv - \epsilon \bar \epsilon\ .
\ee
Notice that there is a subalgebra of two commuting isometries, acting
as shifts on $\sigma$ and $\varphi$, namely $\delta_1=\delta_\alpha $ and 
$\delta_2=-i(\delta_\epsilon -\delta_{\bar \epsilon})$.
The Heisenberg algebra of isometries is thought to persist
at the perturbative level (as discussed in Section 3.1) and hence the 
loop-corrected metric should still have these symmetries.

Finally there is the coordinate system used by Calderbank and Pedersen
\cite{CP}, who classified four-dimensional quaternion-K\"ahler
metrics with two commuting isometries. The metric is given in terms of a 
single function ${\tilde F}(\eta,\rho)$ satisfying a Laplace-like equation
\be\label{CP-laplace}
\rho^2 (\partial_\rho^2 +\partial_\eta^2){\tilde F}(\eta,\rho)=\frac{3}{4}
{\tilde F}(\eta,\rho)\ .
\ee
Using the notation ${\tilde F}_\rho \equiv \partial_\rho {\tilde F}$ and
${\tilde F}_\eta \equiv \partial_\eta {\tilde F}$, we can introduce matrices,
\be\label{QN}
Q=\begin{pmatrix} \ft12 {\tilde F}-\rho {\tilde F}_\rho & -\rho
{\tilde F}_\eta \cr
-\rho {\tilde F}_\eta & \ft12 {\tilde F} + \rho {\tilde F}_\rho
\end{pmatrix}\ ,\qquad
N=\begin{pmatrix} 0 & 1/{\sqrt \rho} \cr  {\sqrt \rho} & \eta/{\sqrt \rho} \end{pmatrix}\ .
\ee
We can now write the metric as \cite{CP}
\be\label{CP-metric}
{\rm d}s^2=\frac{-2\,{\rm det}\,Q}{{\tilde F}^2\rho^2}\,\Big(({\rm d}\rho)^2 +
({\rm d}\eta)^2 \Big)+M^{IJ}(\rho,\eta)\,{\rm d}\phi_I {\rm d}\phi_J\ ,
\ee
where we denote the two extra coordinates by $\phi_I$, which for the universal 
hypermultiplet are related to $\{\varphi,\sigma\}$ from (\ref{real-UHM}). 
The matrix $M$ is related to $Q$ and $N$ by
\be
M^{IJ} = -\frac{2}{{\tilde F}^2\, {\rm det}\,Q}\,\big( N\,Q^2N^t\big)^{IJ}\ .
\ee
Note that (\ref{CP-laplace}) is manifestly invariant under constant
rescalings of ${\tilde F}$, whereas the metric is invariant only after 
rescaling the coordinates $\phi_I$. The overall normalization that 
we have chosen differs from \cite{CP,amtv} by a factor of $2$,
which is convenient for comparison with (\ref{real-UHM}).

The tree-level universal hypermultiplet metric is obtained by taking
\be\label{class-F}
{\tilde F}=\rho^{3/2}\ ,
\ee
which trivially satisfies (\ref{CP-laplace}). The relation between
the Calderbank-Pedersen variables and the ones from (\ref{real-UHM}) is
\be
{\rm e}^{\phi} = \rho^2~ , \qquad \chi = 2
\eta~ , \qquad \s = 2 \phi_1~ , \qquad \varphi=\phi_2~,
\label{CoordRel}
\ee
as can be checked by plugging (\ref{class-F}) into
(\ref{CP-metric}) and changing coordinates as above.

The Calderbank-Pedersen metric has a manifest (and for the case of interest non-compact) 
$U(1)\times U(1)$ isometry group, which acts as shifts on the scalars $\phi_I$ (this is the 
commuting subalgebra of the Heisenberg algebra noted below (\ref{Heis-alg})). Using the 
known Legendre transform techniques, one can dualize these scalars into two tensors. 
The resulting action of such a double tensor multiplet reads
\be \label{CP-DTM-action}
  e^{-1}\mathcal{L}_\mathrm{DT} =
\frac{{\rm det}\,Q}{{\tilde F}^2\rho^2}\,\Big(\partial_\mu \rho\,
\partial^\mu \rho +
\partial_\mu \eta\, \partial^\mu \eta \Big)+ \half M_{IJ}
  H^{\mu I}  H_{\mu}^J\ ,
\ee
where the $H^I$ are a pair of three-form field strengths, $H^{\mu I}=\half
\varepsilon^{\mu\nu\rho\sigma}\partial_\nu B_{\rho\sigma}^I$, and $M_{IJ}$
is the inverse of $M^{IJ}$, $M^{IK}M_{KJ}=\delta^I_J$.

In the coordinates (\ref{CoordRel}), the dual Lagrangian of (\ref{real-UHM})
takes the form \cite{TV1,TV2}
\be \label{DTM-action}
  e^{-1}\mathcal{L}_\mathrm{DT} = - \half\, \p^\mu \phi\, \p_\mu \phi
  - \half\,  \e^{-\phi}\, \p^\mu \chi\, \p_\mu \chi + \half M_{IJ}
  H^{\mu I}  H_{\mu}^J\ ,
\ee
where, after taking into account factors of two from the relation 
$\sigma =2 \phi_1$,
\be
  M_{IJ} = \e^{\phi}\begin{pmatrix} \e^{\phi} + \chi^2
 & - \chi \cr - \chi & 1 \end{pmatrix}\ .
\ee
For reasons that become clear later, we have used a slightly different 
notation than in \cite{TV1,TV2}, in that we have lowered/raised and 
interchanged the labels ``1'' and ``2''.

The two scalars $\phi$ and $\chi$ parameterize the coset 
SL$(2,\fieldR)/\mathrm{O}(2)$. The presence of the tensors breaks 
the SL$(2,\fieldR)$ symmetries to a two-dimensional subgroup generated 
by a certain rescaling of the fields and by the remaining generator of the
Heisenberg algebra (\ref{H-A}) with real parameter $\epsilon=\gamma/2$. 
It acts as a shift on $\chi$ and transforms the tensors linearly into each other \cite{TV1},
\be\label{Heis-symm}
\chi \rightarrow \chi + \gamma\ ,\qquad B^2 \rightarrow B^2 + \gamma B^1\ ,
\ee
with $\phi$ and $B^1$ invariant. Note that the combination 
$\hat{H}^2=H^2-\chi H^1$ is invariant under (\ref{Heis-symm}).

\subsection{Hyperk\"ahler/superspace description}

As quaternion-K\"{a}hler (QK) geometry is rather complicated, the original 
description of hypermultiplet couplings to supergravity in terms of QK geometry
\cite{BW} is not easy to study. As mentioned in the introduction, 
an alternative approach \cite{DWKV,DWRV1,ARV}, based on 
the $N=2$ superconformal tensor calculus \cite{DWLVP}, uses 
the one-to-one correspondence\footnote{This correspondence 
is one-to-one only locally. Globally there may exist a second HKC, 
the double cover of the first one, if a certain topological invariant of the 
quaternion-K\"{a}hler space vanishes \cite{Swann}.} between $4n$-(real) 
dimensional QK spaces and $4(n+1)$-dimensional hyperk\"ahler cones 
(HKC) \cite{Swann,Galicki}. In the superconformal calculus, the HKC is 
realized by introducing an additional hypermultiplet as a compensator. 
As a result the theory is invariant under local $N=2$ superconformal 
transformations, where the combined set of hypermultiplet scalar fields 
(physical and compensator) parametrize the HKC.  If one eliminates the 
compensator by imposing suitable gauge-fixing conditions that fix the 
dilatations and local SU(2) $R$-symmetries, one recovers the matter 
coupled Poincar\'e supergravity. The target space of the 
hypermultiplets of the conformal theory is the HKC, which is a 
hyperk\"ahler space with a conformal homothety and an 
SU(2) isometry that rotates the complex structures. Having described
the theory by a hyperk\"ahler-target sigma-model,
we can use the powerful techniques of projective superspace \cite{proj}.

Projective superspace is a subspace of $N=2$ superspace defined
with the help of an additional complex projective coordinate $\zeta$.
For a recent detailed account of its properties and conventions, we refer to 
Appendix B of \cite{DWRV1}. The $N=2$ projective superfields that we 
need are all power series in the variable $\zeta$ with coefficients that 
project to $N=1$ superfields. More precisely, these are the so-called 
polar and ${\cal O}(2p)$ multiplets \cite{polar,grrwlvu}.

A polar multiplet \cite{polar} has the form
\be
\Upsilon = \sum_{n=0}^{\infty} \Upsilon_n \zeta^n \, .
\ee
As the real structure in projective superspace is given by complex conjugation 
composed with the antipodal map $\zeta^* \rightarrow - 1/\zeta$, the conjugate is
\be
\bar{\Upsilon} = \sum_{n=0}^{\infty} \bar{\Upsilon}_n (- \frac{1}{\zeta})^n \, .
\ee
The polar multiplets are off-shell hypermultiplets; as a 
product of them is again a polar multiplet, they can be used as 
(quaternionic) coordinates which admit a natural gauge action, 
and hence can realize the supersymmetric hyperk\"ahler quotient 
construction of \cite{scal-tens,HKLR}. Superspace actions can be 
written in terms of the
usual $N=1$ measure \cite{polar}
\be
S= \oint_C \frac{{\rm d} \zeta}{2 \pi i \zeta}
\; {\rm d}^4 x \; D^2 \bar{D}^2  G(\uu, \bu, \zeta)\ ,
\label{eq-action}
\ee
where $C$ is a contour in the $\zeta$-plane that generically depends on
the form of the Lagrangian $G$. To make the action invariant under
\emph{local} superconformal symmetries, one simply couples the polar
multiplets to the Weyl multiplet, without changing the form of $G$.
For on-shell hypermultiplets in components, this coupling was described
in \cite{DWKV}.

In \cite{DWRV1}, the hyperk\"ahler cone of
the universal hypermultiplet was described in great detail. For our purposes, the projective superspace formulation in terms of polar multiplets, which makes the SU(1,2) symmetry of the universal hypermultiplet manifest
and linear is most useful; the Lagrange density is given by
\be 
{\hat G}_V={\rm e}^V\bu^i\,\eta_{ij}\uu^j~,~~i=1,2,3
\label{l1}
\ee
where $V$ is a projective superspace $U(1)$ gauge potential\footnote{Although ${\rm e}^V$ acts as a Lagrange multiplier in \eqn{l1}, as explained in \cite{DWRV1}, after reducing to $N=1$ superspace or components, or after duality transformations, this Lagrangian is perfectly sensible.}, and $\eta_{ij}=(-++)$. The factor ${\rm e}^V$ arises because the HKC is the hyperk\"ahler quotient of the $12$-dimensional flat $\s$-model by the diagonal $U(1)$ action:  $\de\uu^i=i\uu^i$. Integrating out $V$ after a certain duality transformation as explained in \cite{DWRV1}, and fixing the gauge $\uu^3=1$ gives the polar multiplet superspace Lagrangian for the hyperk\"ahler cone of the universal hypermultiplet:
\be
G_{\rm HKC}=2\sqrt{-\bu^1\uu^1+\bu^2\uu^2}\ .
\label{GHKCUH}
\ee
Notice that the quotient Lagrangian has only a
manifest U(1,1) rather than SU(1,2) invariance.

The other multiplet of interest, the ${\cal O}(2p)$ multiplet, is real and of the form
\be
\eta^{(2p)} (\zeta) = \frac{1}{\zeta^p} \sum_{n=0}^{2p} \eta_n^{(2p)} \zeta^n \, .
\ee
Particularly important for us will be the ${\cal O}(2)$ multiplet:
\be
\eta^{(2)} = \frac{\bar{z}}{\zeta} + x - z \zeta \, ,\label{eta2}
\ee
where $z$ projects to an $N=1$ chiral superfield and $x$ projects to a real linear superfield. 

In Section 4, we will also need the ${\cal O}(4)$ multiplet:
\be
\eta^{(4)} = \frac{\bar{z}}{\zeta^2} + \frac{\bar{v}}{\zeta} + x - v\zeta + 
z \zeta^2 \, ,
\ee
where $z$ is constrained as for the ${\cal O}(2)$ multiplet, $v$ is a complex
linear superfield which has one complex physical scalar, and $x$ projects
to an auxiliary real unconstrained $N=1$ superfield. In total, the 
${\cal O}(4)$ multiplet has four real physical degrees of freedom.

The ${\cal O}(2)$ multiplet is precisely the $N=2$ tensor multiplet. Because
dualizing scalars into tensors with respect to commuting isometries of a 
quaternionic geometry in Poincar\'e supergravity translates to dualizing with 
respect to triholomorphic isometries of the corresponding hyperk\"{a}hler cone, 
the double tensor multiplet description of the universal hypermultiplet is equivalent
to a description of the superconformal universal hypermultiplet (obtained by 
adding a compensating hypermultiplet) in terms of two $N=2$ tensor multiplets 
$\eta_1$ and $\eta_2$.

In general, an HKC may have several different pairs of commuting isometries. 
For the case of the universal hypermultiplet, as explained in \cite{DWRV1}, 
there are at least three choices, yielding three tensor multiplet descriptions. The particular choice we study here is the one that corresponds to the HKC-lift of the $u(1)\times u(1)$
subalgebra inside the Heisenberg algebra \eqref{H-A} (a second choice is described in Appendix B). 
In \cite{DWRV2}, the relation between isometries on a quaternion-K\"ahler 
space Q and their lift to triholomorphic isometries on the HKC above Q 
was written down explicitly. Using the relation between the $S,C$ coordinates 
of \eqref{S-C} and the HKC variables given in \cite{DWRV1}, the procedure 
given in \cite{DWRV2} yields the action of the Heisenberg subalgebra on 
the fields $\uu^i$. We give more details on the derivation of the generators 
in Appendix A. The result is:
\be \label{GenP}
T_1=  \left(
\begin{array}{rrr}
1 & -1& ~0\\ 1&-1&~0\\0&0&~0
\end{array}\right)~~,~~~~
T_2=  \left(
\begin{array}{rrr}
0 & ~0& 1\\0&0&1\\-1&1&~0
\end{array}\right)~~,~~~~
T_3=  \left(
\begin{array}{rrr}
~0 & ~0&i\\0&0&i\\i&-i&~0
\end{array}\right)~~.
\ee
By exponentiating these generators, one finds the linear action of the
Heisenberg group on the polar multiplets $\uu^i$. The correspondence 
with \eqref{H-A} is such that $T_1$ and $T_2$ generate the shifts in 
$\sigma$ and $\varphi$ respectively. These generators obey the algebra
\be
[T_2,T_3]=2iT_1~,
\ee
with all other commutators vanishing. Note that the above matrix representation 
of the algebra obeys a much stronger relation: $T_1T_m=T_mT_1=0$.

The dualization with respect to $T_1$ and $T_2$ in superspace was worked out in detail in \cite{DWRV1}.\footnote{More precisely, the generators considered in that work are the transpose of $T_1$ and $T_2$ in (\ref{GenP}). However this difference does not affect the end product of the dualization.} The resulting action has two $N=2$ tensor multiplets and the superspace Lagrange density is\footnote{For later convenience, we have interchanged 
the indices $1$ and $2$ of $\eta_I$ and $T_I$ in (\ref{GenP}), as well 
as the roles of $T_I$ and $T'_I$, relative to the conventions used in \cite{DWRV1}.}
\be\label{class-G-eta}
G(\eta_1,\eta_2)=\frac{(\eta_2)^2}{\eta_1}\ .
\ee
Notice that this is a function homogeneous of first degree and there is
no explicit $\zeta$ dependence. As explained in \cite{DWRV1} this is
a consequence of the superconformal invariance. The Lagrange density \eqref{class-G-eta} is gauge-equivalent to the double-tensor multiplet Lagrangian \eqref{DTM-action}. The correspondence between the HKC and QK formulations is discussed further in the next subsection.

\subsection{Relation between the HKC and QK metrics}

In this subsection, we explain in more detail how the HKC and QK metrics are related. As known since \cite{gh,HKLR,PP}, the metric on a $4n$-dimensional hyperk\"{a}hler space with
$n$ commuting isometries can be written in the form
\be
{\rm d}s^2 = U_{IJ}(r)\, {\rm d}\vec{r}^{\, I} \cdot {\rm d}\vec{r}^{\, J} +
[U^{-1}]^{IJ} (r)\, ({\rm d}t_I +
 A_I) ({\rm d}t_J + A_J) \, ; \qquad I,J = 1, ... , n \, , \label{metric}
\ee
where the coordinates $t_I$ parametrize the $n$ U(1) isometries and
$\vec{r}^{\, I}$ are $n$ three-dimensional vectors which are the moment maps associated with $\partial_{t_I}$. Finally, $A_I \equiv {\vec A_{IJ}}\cdot {\rm d}
{\vec r}^J$ are one-forms determined by the Bogomol'nyi equation
\be \label{Bog-eqn}
{\cal R}_{r_a^I r_b^J}^{(K)} = \sum_{c} \varepsilon_{abc} \nabla_{r_c^I} U_{JK} \,\,\, , \qquad a,b,c = 1,2,3 \,\,\, ,
\ee
where ${\cal R}^{(K)}$ is the SU(2) curvature (field strength) of $A_K$.

An explicit solution to (\ref{Bog-eqn}) was found in \cite{scal-tens,proj,HKLR} by
regarding the manifold as the target space of a supersymmetric nonlinear $\sigma$-model.
Then the metric (\ref{metric}) is encoded in a single function $F$ 
depending on $3n$ variables $x^I,z^I,{\bar z}^I$. 
Supersymmetry requires this function to satisfy a Laplace-like equation
\be \label{constr}
F_{x^Ix^J}+F_{z^I{\bar z}^J}=0\ ,
\ee
where $F_{x^Ix^J}$ denote the second derivatives. The constraint (\ref{constr}) can be solved in terms of the following contour integral representation \cite{proj,HKLR}:
\be \label{F-cont}
F = Re \oint \frac{{\rm d} \zeta}{2 \pi i \zeta} G(\eta^I(\zeta), \zeta)\ ,
\qquad \eta^I =
\frac{\bar{z}^I}{\zeta} + x^I - z^I \zeta \, .
\ee
The hyperk\"ahler metric \eqref{metric} is then determined by
\be
U_{IJ} = -\frac{1}{2} F_{x^I x^J}\ , \qquad   A_I = i (F_{z^K x^I}
 dz^K - F_{\bar{z}^K x^I} d\bar{z}^K) \, , \label{M}
\ee
where
\be
x^I = r^I_3 \,  \qquad z^I = \frac{r^I_1 + i r^I_2}{2} \, , \qquad \bar{z}^I
= \frac{r^I_1 - i r^I_2}{2} \ . \label{XZR}
\ee

The scalars of the universal hypermultiplet parametrize a four-dimensional quaternion-K\"ahler space corresponding to an eight-dimensional HKC. As discussed above, four-dimensional QK
manifolds with two commuting isometries were classified by
Calderbank and Pedersen in terms of a single function ${\tilde F}$ satisfying the Laplace-like equation \eqref{CP-laplace}. Given a solution ${\tilde F}$ one can construct the hyperk\"ahler cone metric \eqref{metric} by 
writing \cite{CP}
\be
U_{IJ} = \frac{\tilde{F}\det{Q}}{|q|^2} \left[(N^{-1})^t \,Q^{-1} N^{-1}\right]_{IJ}~,
\label{CP}
\ee
where $Q$ and $N$ are defined in (\ref{QN}), and the 8 real coordinates of the HKC
are the QK coordinates 
$\rho,\eta,t_I\equiv\phi_I, \,I=1,2$ along with the 4 real 
components of the quaternion $q$ (the hypermultiplet 
compensator). The one forms $A_I$ can be computed by solving (\ref{Bog-eqn}),
or equivalently, by finding $F$ and using (\ref{M}); see also \cite{SZ}.

The precise relations between the QK and HKC variables are further
given by \cite{CP}
\be
\rho = \frac{|\vec{r}^{\, 1} \times \vec{r}^{\, 2}|}{|\vec{r}^{\, 1}|^2}
\ , \qquad \eta = \frac{\vec{r}^{\, 1} \cdot \vec{r}^{\, 2}}
{|\vec{r}^{\, 1}|^2} \ , \label{RE}
\ee
and (correcting a misprint in \cite{CP}, which was already noted
in \cite{BCGP})
\be \label{Rq}
|\vec{r}^{\, 1}
\times \vec{r}^{\, 2}| = |q|^4 / \tilde{F}^2 \ .
\ee
Notice that all these relations are consistent with the scaling weights
of the coordinates of the HKC, whereby we assign weights 2,1 and 0 to the
coordinates $\vec{r}^{\, I}, q$ and $t_I$ respectively. This assignment can be understood as follows: A generic cone metric can be put in the form ${\rm d}s^2 = {\rm d}r^2 + r^2 {\rm d}\Omega^2$, where $r$ is the radial coordinate and ${\rm d}\Omega^2$ is the metric on the base space. The homothety that this metric has keeps ${\rm d}\Omega^2$ invariant and in four dimensions\footnote{We note that as this action is dimension dependent so are the resulting scaling weights. For a general discussion and also the weigths of the fields in various multiplets in $d=5$ see \cite{BCDWGHVVP}.}  acts on $r$ as $r \rightarrow \alpha r$. Clearly that means that under the homothety ${\rm d}s^2 \rightarrow \alpha^2 {\rm d}s^2$ or in other words that the metric has scaling weight 2. Now, looking at (\ref{metric}) and (\ref{CP}) and keeping in mind that the QK quantities $\rho$, $\eta$ and $t_I$ have to be inert under this rescaling, we deduce the above weights of $\vec{r}^{\, I}$ and $q$.

To recapitulate: all of the above implies that comparison between the 
descriptions in terms of the
functions $F$ and $\tilde{F}$ amounts to comparison between the matrices
$U_{IJ}$ in (\ref{M}) and (\ref{CP}), possibly up to a coordinate
transformation which is equivalent to taking two linearly independent 
constant-coefficient combinations of the two commuting isometries as the 
new isometries.

For completeness, before concluding this section we comment on the lifting of the symmetry (\ref{Heis-symm}) to the tensor multiplet description in projective 
superspace. This is also helpful for identifying unambiguously the component-field content of the two $\eta_I$ tensor multiplets. Combining (\ref{CoordRel}), (\ref{RE}) and the invariance of $\phi$ (equivalently, $\rho$), we see that the residual Heisenberg transformation acts as
\be
\vec{r}^{\, 2} \rightarrow \vec{r}^{\, 2} + \frac{\gamma}{2} \vec{r}^{\, 1} \, , \qquad \vec{r}^{\, 1} \rightarrow \vec{r}^{\, 1} \, .
\ee
Hence from (\ref{F-cont}), (\ref{XZR}) it follows that
\be \label{Hs_eta}
\eta_2\rightarrow \eta_2 + \frac{\gamma}{2} \eta_1 \ ,\qquad \eta_1\rightarrow \eta_1\ .
\ee
So indeed $\eta_1$ contains $\phi$ and $B_1$ whereas $\eta_2$ contains $\chi$ and $B_2$. Finally, note that the action determined by (\ref{class-G-eta}) is invariant under the residual Heisenberg symmetry (\ref{Hs_eta}) as an infinitesimal transformation on $G(\eta_1,\eta_2)$ yields a term linear in $\eta_2$, which vanishes under the superspace integral.

\section{Perturbative corrections}
\setcounter{equation}{0}

\subsection{One-loop metrics}

The classical action of the universal hypermultiplet is invariant under eight symmetry transformations whose explicit form was found in \cite{dWPP}. While four of them are not expected to survive beyond the classical limit, the other four are thought to play a role at the quantum level. One of them, the classical symmetry acting as an $SO(2)$ transformation on the RR scalars $\chi$
and $\varphi$ or, equivalently, on $\eta$ and $\phi_2$ (see (\ref{CoordRel})):
\be \label{extra-U1}
C \rightarrow e^{i \theta} C \, , \qquad S \rightarrow S
\ee
with $\theta$ real, reduces in the full quantum 
theory to \cite{BB}
\be
Re C \rightarrow Im C \, , \qquad Im C \rightarrow - Re C \, ,
\ee
i.e. to the case $\theta = \frac{\pi}{2}$. 

The remaining three are 
the transformations (\ref{HeisTr}). They are believed to be preserved by perturbative corrections \cite{S}. It was noticed in \cite{BB} that they 
satisfy a Heisenberg algebra (in our notation: (\ref{Heis-alg})). It was 
further argued there that membrane and fivebrane instanton corrections lead 
to discrete identifications due to the charge quantization conditions that 
must be obeyed by the corresponding field strengths. Hence the continuous 
Heisenberg symmetry is broken non-perturbatively to a discrete subgroup.

The perturbative corrections to the moduli space of the universal
hypermultiplet were first studied in \cite{S} with the conclusions
that they arise only at one-loop and that they can be absorbed in
a field-redefinition. However, recently it was found that the second conclusion is incorrect \cite{amtv}. The authors of \cite{amtv}
performed explicit stringy one-loop computation of the three-point
amplitude with external graviton/NS-NS tensor and two RR scalars
and showed that this induces a genuine (unremovable) deformation
of the hypermultiplet target space. This deformation is still
invariant under the Heisenberg algebra of isometries, and in particular under 
the shift of the coordinate $\eta$ (not to be confused with the
projective superfields $\eta_I$). However, it does not preserve the K\"{a}hler
structure of the classical moduli space. It is encoded in the
following change of (\ref{class-F}): 
\be \label{quant-F} 
\tilde{F}= \rho^{3/2} -\hat{\chi} \rho^{-1/2} \, , 
\ee
where $\hat{\chi}$ is a constant parameter. 
Moreover, as shown in \cite{amtv}, quantum corrections proportional to $\hat \chi$ also modify 
the relation between the Calderbank-Pedersen coordinate $\rho$ and the 
four-dimensional dilaton $\phi$. We will see in the following how to
recover (\ref{quant-F}) from the perspective of projective superspace.

\subsection{Superspace derivation}

In this subsection we write the most general deformation of the HKC
corresponding to the universal hypermultiplet compatible with the
Heisenberg algebra, and find the dual description in terms of
${\cal O}(2)$ multiplets (\ie, $N=2$ tensor multiplets). In the next subsection we show that it indeed reproduces the one-loop
deformation (\ref{quant-F}) of \cite{amtv}.

As derived at the end of the previous section, the Heisenberg algebra
in projective superspace acts as (\ref{Hs_eta}). We should therefore
look for a function $G(\eta_1,\eta_2)$ which is 
homogeneous of first degree (due to the conformal symmetry on the HKC), and
invariant under (\ref{Hs_eta}). The classical part is given by 
(\ref{class-G-eta}). The deformation describing the one-loop correction should
be proportional to $\eta_1$ as this is the multiplet that contains the dilaton. Up to
a constant parameter $\hat \chi$, these constraints are solved by
\be \label{def-G}
G(\eta_1,\eta_2)=\frac{(\eta_2)^2}{\eta_1}-2\hc \, \eta_1\ln\eta_1 ~ .
\ee

We now derive this result from the hyperk\"ahler quotient
construction on the twelve-dimensional flat hyperk\"ahler space on
which the isometries act linearly. Classically, this quotient was
obtained from the gauged Lagrangian density (\ref{l1}).
It turns out that it is a bit
tricky to obtain a deformed HKC metric. What first comes to mind
is to deform the flat metric of the initial $\sigma$-model whose
hyperk\"{a}hler quotient is the HKC. Looking for the most general
metric preserved by the Heisenberg algebra: \be
g_{ij}(T_m)^j{}_k=(T^\dag_m)_i{}^jg_{jk}~, \ee we find \be
g_{ij}\propto(\eta_{ij}-\ep\eta_{ik}(T_1)^k{}_j)~. \label{g} \ee
However, substituting $\eta\to g$ in \eqn{l1} with a constant
deformation $\ep$ does not modify the universal hypermultiplet, as
the deformation can be absorbed by a linear change of variables
acting on $\uu^i$. Fortunately, there is another way to proceed.
Namely, we deform the $U(1)$ gauge group with respect to which we
take the quotient by adding a term proportional to
$T_1$.\footnote{Strictly speaking, this changes $U(1)$ to a
noncompact gauge symmetry but we have not investigated the
consequences of this.} Because $(T_1)^2=0$, this has the effect of
deforming the metric as in \eqn{g} with $\ep\propto V$. Thus the
gauged Lagrange density is now: 
\be 
{\hat G}_V = {\rm e}^V\bu^i(\eta_{ij}-4\hc
V\eta_{ik}(T_1)^k{}_j)\uu^j~, \label{l2} 
\ee 
where $\hc$ is a constant deformation parameter.

We now dualize with respect to two $U(1)$ isometries that commute with and 
are independent of the gauge symmetry; again, there are a number of 
choices,\footnote{An alternative
dualization in terms of two different generators, one of which corresponds to the symmetry (\ref{extra-U1}), is given in Appendix B.} but as explained for the classical Lagrangian, we 
focus on the symmetries generated by $T_1$ and $T_2$.
We use the explicit relations $(T_2)^2=-T_1$ and $T_1T_m=T_mT_1=0$, to find
\be
{\rm e}^{V_1T_1+V_2T_2}=1+(V_1-\ft12(V_2)^2)T_1+V_2T_2~.
\ee
This implies that we may write down the first-order Lagrange density
\be
{\rm e}^V\bu^i[\eta_{ij}+\eta_{ik}(V_1-\ft12(V_2)^2-4\hc V)(T_1)^k{}_j+
V_2(T_2)^k{}_j)]\uu^j-\eta_1V_1-\eta_2V_2~,
\label{l3}
\ee
where $\eta_I,~ I=1,2$ are $\co(2)$ multiplets; if we integrate out
$\eta_I$, we find that both $V_I$ are pure gauge and recover \eqn{l2}. On the other hand, if we integrate out $V_I$ and $V$, after dropping total derivative terms, we get the dual theory expressed purely in terms of $\eta_I$. Before we begin this calculation, we simplify \eqn{l3} by shifting $V_1$ and obtain:
\be
{\rm e}^V(X_0+V_1X_1+V_2X_2)-\eta_1(V_1+\ft12(V_2)^2+4\hc V)-\eta_2V_2~,
\label{l4}
\ee
where
\begin{eqnarray}
X_0&\equiv&\bu^i\eta_{ij}\uu^j=-\bu^1\uu^1+\bu^2\uu^2+\bu^3\uu^3~,
\nonumber\\
X_1&\equiv&\bu^i\eta_{ik}(T_1)^k{}_j\uu^j=-(\bu^1-\bu^2)(\uu^1-\uu^2)~,
\nonumber\\
X_2&\equiv&\bu^i\eta_{ik}(T_2)^k{}_j\uu^j=-\bu^3(\uu^1-\uu^2)-(\bu^1-\bu^2)
\uu^3~.
\end{eqnarray}
Varying \eqn{l4} with respect to $V,V_1,$ and $V_2$ respectively, we find
\be
{\rm e}^V(X_0+V_1X_1+V_2X_2)=4\hc\eta_1~,~~
{\rm e}^VX_1=\eta_1~,~~
{\rm e}^VX_2=\eta_2+V_2\eta_1~.
\ee
We solve these for $V,V_1,$ and $V_2$:
$$
V=\ln\eta_1-\ln X_1~~,~~~V_2=-\frac{\eta_2}{\eta_1}+\frac{X_2}{X_1}~~,
$$
\be V_1=4\hc-\frac{X_0}{X_1}+\frac{\eta_2X_2}{\eta_1X_1}
-\left(\frac{X_2}{X_1}\right)^2~~. 
\ee 
Substituting back into \eqn{l4}, and dropping numerous total derivatives 
of the form $\eta(\bu+\uu)$, we find\footnote{In particular, $X_2/X_1$
and $X_0/X_1+\ft12(X_2/X_1)^2$ both have the form $\bu+\uu$.} \be
\ft12\frac{(\eta_2)^2}{\eta_1}-4\hc \, \eta_1\ln\eta_1 \, . 
\ee

However, recall that for convenience we have redefined the canonical 
generator in the direction of $t_1$ (or equivalently, $\sigma$) by a 
factor of two (see (\ref{Gen})). So the dual variable $\eta_1$ is twice 
bigger than it should be. Rescaling $\eta_1 \rightarrow \eta_1/2$, and 
dropping a term linear in $\eta_1$ (which vanishes under the superspace 
integral) we obtain our final result
\be
\frac{(\eta_2)^2}{\eta_1}-2\hc \, \eta_1\ln\eta_1 \, , \label{leta}
\ee
as stated in the beginning of this subsection.

\subsection{Comparing the one-loop metric and superspace action}

In this subsection we will show that the superspace Lagrange density (\ref{leta}) gives exactly the metric determined by (\ref{quant-F}). We thereby make use of the general results of Section 2.3.

The matrix $U_{IJ}$ of (\ref{CP}) with $\tilde{F}$ as in (\ref{quant-F}) has the form
\be \label{Mtilde}
U_{IJ}  = \frac{1}{|\vec{r}^{\, 1}|} \left(\begin{array}{cc} (2\eta^2 - \rho^2 - \hat{\chi}) & - 2 \eta \\ - 2 \eta & 2\end{array} \right) ~,
\ee
where we have used (\ref{RE}) and (\ref{Rq}).

Now let us find the matrix $-\frac{1}{2} F_{x^I x^J}$ with $I,J =1,2$ (see (\ref{M})). The deformation piece in (\ref{leta}) is
\be
G^{def} = - 2\hat{\chi} \, \eta_1 \ln \eta_1 \qquad \Rightarrow \qquad
F_{x^1 x^1}^{def} = \frac{2\hat{\chi}}{\sqrt{(x^1)^2 + 4 z^1 \bar{z}^1}} =
\frac{2\hat{\chi}}{|\vec{r}^{\, 1}|} \, .
\ee
For the classical contribution we find:
\bea \label{f}
\hspace*{-1cm} F^{cl} &=& \oint \frac{d \zeta}{2 \pi i \zeta} \,\, \frac{\eta_2^2}{\eta_1} = \frac{(x^2)^2}{|\vec{r}^{\, 1}|}  \\ \nn \\ &+& \frac{((x^1)^2+2z^1\bar{z}^1) ((z^1 \bar{z}^2)^2+(z^2 \bar{z}^1)^2) - 2z^1 \bar{z}^1 (x^1x^2(z^2\bar{z}^1+z^1\bar{z}^2)-2z^1\bar{z}^1z^2\bar{z}^2)}{2 (z^1 \bar{z}^1)^2 |\vec{r}^{\, 1}|} \, , \nn
\eea
where the contour of integration is the same as the one in Figure 1 of \cite{HKLR}. Despite its unappealing form the last result gives the following elegant expressions for the second derivatives of interest:
\bea
F^{cl}_{x^2 x^2} = - \frac{4}{|\vec{r}^{\, 1}|} \, , \qquad F^{cl}_{x^1 x^2} = \frac{4 \vec{r}^{\, 1} \cdot \vec{r}^{\, 2}}{|\vec{r}^{\, 1}|^3} = \frac{4 \eta}{|\vec{r}^{\, 1}|} \, , \nn
\eea
\be \label{Der}
F^{cl}_{x^1 x^1} = -4 \frac{[(\vec{r}^{\, 1} \cdot \vec{r}^{\, 2})^2 - \frac{1}{2} (\vec{r}^{\, 1} \times \vec{r}^{\, 2})^2]}{|\vec{r}^{\, 1}|^5} = - \frac{4}{|\vec{r}^{\, 1}|} \left( \eta^2 - \frac{\rho^2}{2} \right) \, ,
\ee
where we have used (\ref{RE}). Hence we obtain the complete result
\be
- \frac{1}{2} F_{x^Ix^J} = \frac{1}{|\vec{r}^{\, 1}|} \left(\begin{array}{cc} (2\eta^2 - \rho^2) & - 2\eta \\ - 2\eta & 2\end{array} \right) - \frac{\hat{\chi}}{|\vec{r}^{\, 1}|} \left(\begin{array}{cc} 1 & 0 \\ 0 & 0\end{array} \right) \, ,
\ee
which agrees precisely with (\ref{Mtilde}).

\section{Proposal for instanton corrections}
\setcounter{equation}{0}

On general grounds the perturbative hypermultiplet moduli space metric is expected to be corrected by instanton effects. As already recalled, these are due to membranes and five-branes wrapping supersymmetric cycles in the internal space. The relevant instanton actions were calculated in \cite{BB, GS, TV1}. However, these actions are only the first step in the determination of the quantum corrected moduli space metric. The main obstacle for performing explicitly such a computation is the fact that in string/M theory, unlike in second quantized field theory, the rules for calculating the one-loop fluctuation determinants have not yet been established.

Addressing this problem is beyond the scope of our paper. In the
current section we have a more modest goal. Namely, we make a
conjecture for the form of the nonperturbative universal
hypermultiplet moduli space metric due to five-brane instantons.\footnote{Explicit five-brane instanton calculations in the semiclassical and supergravity approximation 
are performed in \cite{DTV}.}
An obvious virtue of our
proposal is that it reduces to the perturbative result in the
asymptotic region. Further investigation of the proposed metric is
more than desirable (although technically not easy) and we hope to
report on this in the near future.\footnote{We note that quantum corrected metrics were also proposed in
\cite{SK}. The author considered deformations of the QK
metric (rather than the corresponding HKC one) for the case of 
vanishing five-brane charge ({\it i.e.} only $2$-brane instantons present).}

To explain the rationale behind our proposal we now recall the
projective superspace description of the four-dimensional hyperk\"ahler 
moduli space metric on the Coulomb branch of pure three-dimensional $N=4$ 
Yang-Mills theory with gauge group $SU(2)$. 
The perturbative (one-loop corrected) moduli space metric is 
the Taub-NUT one (with mass parameter $m=-1$). Taking into account the instanton effects it becomes the 
Atiyah-Hitchin metric \cite{SW,DKMTV}. These two metrics are determined, in 
the language of Sections 2.2 and 2.3, by a single superfield with 
\be F_{TN} = \oint_{C_0} \, \frac{d\zeta}{2\pi i \zeta} \,\, (\eta^{(2)})^2 
+ m \oint_{C} \, \frac{d\zeta}{\zeta} \,\, \eta^{(2)} (\ln{\eta^{(2)}} - 1) \ ,
\label{FTNUT} \ee 
and 
\be
F_{AH} =  \oint_{C_0} \, \frac{d\zeta}{2\pi i \zeta} \,\, 
\eta^{(4)} - \oint_{C^{\prime}} \, \frac{d\zeta}{\zeta} \,\, \sqrt{\eta^{(4)}}\ , \label{FAH} 
\ee 
respectively \cite{IR} (notice that our normalization of the $\eta$'s is different from the one in \cite{IR}). The contour $C_0$ encloses the origin and $C$ ($C^{\prime}$) - all two (four) roots of $\eta^{(2)}$ ($\eta^{(4)}$).\footnote{For more details on the contours see \cite{CRW}, page 4.} Asymptotically $\eta^{(4)} \rightarrow (\eta^{(2)})^2$ \cite{CRW} and so the Atiyah-Hitchin metric tends to the Taub-NUT one (with $m=-1$) as is known to be the case.

From (\ref{FTNUT}) and (\ref{FAH}) we can see that the full moduli space Lagrange density has the same form as the perturbative one, but is written in terms of an $\co(4)$ multiplet rather than an $\co(2)$ one.\footnote{The $\log$ term in (\ref{FTNUT}) seems to defy this rule, but we recall that it is present only to define the correct contour of integration. In (\ref{FAH}) this goal is already achieved by the square root.} It is tempting to try the same trick for the universal hypermultiplet; there are two $\co(2)$ mulitplets $\eta_I$ in \eqn{leta}, but any Lagrange density involving an $\co(4)$ multiplet $\eta^{(4)}$ must have at least two terms involving $\eta^{(4)}$, as otherwise the auxiliary $N=1$ superfield $x$ contained in $\eta^{(4)}$ cannot be consistently eliminated; thus the only candidate for ``promotion'' to $\eta^{(4)}$ is $\eta_1$. Moreover, 
conformal symmetry requires no explicit $\zeta$-dependence in the Lagrange density, 
and scales an ${\cal O}(4)$ multiplet with weight 2. Hence our proposal is
\be \label{F_UH}
F_{UH} = \oint\frac{d\zeta}{2\pi i \zeta}\left(\frac{(\eta_2)^2}{\sqrt{\eta^{(4)}}}-2\hc\sqrt{\eta^{(4)}}\right)~.
\ee
This function determines the K\"{a}hler potential of the HKC,  associated to the universal hypermultiplet moduli space, via the generalized Legendre transform of \cite{polar}.

The reason why (\ref{F_UH}) is suitable to describe the nonvanishing five-brane charge instantons is due to the symmetries that it preserves. Recall that there are three charges that a general instanton can carry, one of which descends from the $5$-brane charge of M-theory and the other two from the M$2$-brane charge (due to the existence of two $3$-cycle homology classes) \cite{BB}. These charges $Q_i$, $i=1,2,3$ are related to the Noether currents corresponding to the three isometries generated by our $T_1, T_2, T_3$. Now, the $5$-brane charge is associated precisely to $T_1$ and hence breaking that isometry (which is exactly what happens as a result of the substitution $\eta_1 \rightarrow \sqrt{\eta^{(4)}}$) signals the presence of five-brane instantons. On the other hand, due to the ${\cal O} (2)$ multiplet $\eta_2$ in (\ref{F_UH}), $F_{U\!H}$ still has the isometry generated by $T_2$ and therefore does not include contributions from 2-brane instantons carrying the associated charge $Q_2$. Finally, going to the description in terms of one ${\cal O} (2)$ and one ${\cal O} (4)$ multiplet breaks also the isometry generated by $T_3$ (which at the perturbative level intertwines the two ${\cal O} (2)$ multiplets; see (\ref{Hs_eta})). Thus our nonperturbative proposal also includes the contribution of $2$-brane instantons with $Q_3$ charge. Summarizing, we propose that (\ref{F_UH}) describes the metric for the case of instanton corrections with two nonvanishing charges: $(Q_1, Q_3)$.\footnote{Note that this is consistent with the explicit supergravity calculation of \cite{TV1} of the five-brane instanton action, where it was found that the action contains not only a term proportional to the $5$-brane charge $Q_1$ but also another one proportional to the asymptotic value of the field $\chi$, which is the representation of the third charge in the double-tensor formulation.}
We are currently investigating the explicit form and properties of the QK metric corresponding to $F_{U\!H}$.

The fully corrected universal hypermultiplet including {\em all} instanton corrections must also break the symmetry that comes from the remaining tensor multiplet $\eta_2$. It is not clear how to implement this. Naively promoting $\eta_2\to\sqrt{\eta_2^{(4)}}$ clearly cannot work. However, it is possible that one needs to choose a different basis than $\eta_{1,2}$ before changing the superfields; further, the nonperturbative action may involve other forms of the hypermultiplet than just ${\cal O} (4)$, \eg, $\eta_2\to(\eta_2^{(6)})^{(1/3)}$, etc. Clearly, some new insights will be needed to make progress on this problem.

\section*{Acknowledgements}
This project was initiated during the Simons Workshop in Mathematics and Physics, Stony Brook
2003. We have benefited from discussions with many of the participants.
S.V. also thanks B. de Wit and P. Vanhove for stimulating discussions.
The research of L.A. was supported in part by DOE grant 
DE-FG$02$-$95$ER$40899$. The research of M.R. was 
supported in part by NSF Grant No. PHY-0098527.

\appendix
\section{Lifting the Heisenberg algebra to the HKC}
\setcounter{equation}{0}

In accord with \cite{DWRV1,DWRV2}, we denote the 
coordinates of the eight-dimensional HKC corresponding to the 
universal hypermultiplet by $z_{+}^I$ and $z_{-I}$, $I=1,2$, and 
their complex conjugates. Descending to the underlying QK 
space can be achieved by introducing\footnote{We follow the 
notation of \cite{DWRV2} for convenience. In \cite{DWRV1} $w$ is denoted by $u$.}:
\be
z_{-2} = {\rm e}^{2z} \, , \qquad z_{-1} = {\rm e}^{2z} w \, , \qquad z_{+}^2 = \frac{\xi}{2} \, , \qquad z_{+}^1 = v\ ,
\ee
and taking $\xi = 0$. The coordinate $z$ scales out, and the
resulting four-dimensional QK metric is determined by the 
K\"{a}hler potential \cite{DWRV1}
\be \label{Kal}
K = \ln \, (1 - w^{\prime} \bar{w}^{\prime} - v^{\prime} \bar{v}^{\prime}) \, ,
\ee
where
\be \label{wpvp-wv}
w^{\prime} = w (1 - v \bar{v}) \qquad {\rm and} \qquad v^{\prime} = \bar{v} \, .
\ee
From (4.14) of \cite{DWRV2} we see that the triholomorphic isometries of the HKC induce the following transformations of the QK coordinates:
\bea \label{var1}
\delta w &=& i (T^2{}_2 - T^1{}_1) w + i T^3{}_1 w v + i w^2 (T^1{}_2 - v T^3{}_2) - i \bar{v} \left(\frac{1}{1-v \bar{v}} -w \bar{w}\right) T^2{}_3 \nn \\
&+& i v \bar{v} w \bar{w} T^2{}_1 - \frac{i}{1 - v \bar{v}} T^2{}_1 \, , \nn \\ \nn \\
\delta v &=& i (T^1{}_1 - v T^3{}_1) v + i T^1{}_3 - i v T^3{}_3 + i \bar{w} (1 - v \bar{v}) (T^2{}_3 + v T^2{}_1) \, ,
\eea
where $T^m{}_n$, $m,n = 1,2,3$ are the matrix elements of the symmetry generators.

We want to find out which generators correspond to the transformations in (\ref{HeisTr}). For that purpose we note that the relationship between the metrics determined by the K\"{a}hler potentials in (\ref{S-C}) and (\ref{Kal}) is given by
\be \label{wpvp-CS}
w^{\prime} = \frac{1-S}{1+S} \qquad {\rm and} \qquad v^{\prime} = \frac{2C}{1+S} \, .
\ee
Hence combining (\ref{wpvp-wv}), (\ref{wpvp-CS}) and \, $\delta S = i \alpha + 2 \bar{\epsilon} C$, \, $\delta C = \epsilon$ \, we obtain
\bea \label{var2}
\delta w &=& \epsilon (1+w) w v - \bar{\epsilon} \bar{v} \left( \frac{1}{1 - v \bar{v}} - w \bar{w} \right) - \frac{i \alpha}{2} \left( 2 w + w^2 - w \bar{w} v \bar{v} + \frac{1}{1 - v \bar{v}} \right) \nn \\ \nn \\
\delta v &=& - \epsilon v^2 + \bar{\epsilon} [1 + \bar{w} (1 - v \bar{v})] + \frac{i \alpha}{2} v [1 + \bar{w} (1 - v \bar{v})] \, .
\eea
Comparison of (\ref{var1}) and (\ref{var2}) yields:
\be
T_{\epsilon}=  \left(
\begin{array}{rrr}
0 & ~0& ~0\\ 0& ~0&~0\\ -i& ~i& ~0
\end{array}\right)~~,~~~~
T_{\bar{\epsilon}}=  \left(
\begin{array}{rrr}
~0 & ~0& -i\\ ~0& ~0& -i\\ ~0& ~0& ~0
\end{array}\right)~~,~~~~
T_{\alpha}=  \left(
\begin{array}{rrr}
\frac{1}{2} & - \frac{1}{2}& ~0 \\ \frac{1}{2} & - \frac{1}{2}& ~0 \\ 0& 0& ~0
\end{array}\right)~~.
\ee
It will be convenient for us to define
\be \label{Gen}
T_1 = 2 T_{\alpha} \, , \qquad T_2 = - i (T_{\epsilon} - T_{\bar{\epsilon}}) \, , \qquad T_3 = - (T_{\epsilon} + T_{\bar{\epsilon}}).
\ee
Clearly, as mentioned in the main text, $T_1$ and $T_2$ generate shifts along $\sigma$ and $\varphi$ respectively.

\section{An alternative dualization}
\setcounter{equation}{0}

An alternative dualization is also possible with respect to the two $U(1)$ isometries generated by $T_1$ and
\be \label{T_2^p}
T'_2= \frac13 \left(
\begin{array}{rrr}
1 & ~0& 0\\ ~0&1&0\\0&0&-2
\end{array}\right)~~.
\ee
This is of interest as, following the same procedure as in the previous Appendix, one can verify that the generator $T_2'$  induces precisely the symmetry \eqn{extra-U1}. Since the latter can survive when only five-brane instanton contributions are taken into account (which means that only the isometry generated by $T_1$ is broken whereas the other Heisenberg symmetries are still continuous, as recalled in Section 4), it may be useful for some purposes to have a formulation in which this symmetry is explicit.

As $T_2'$ commutes with $T_1$, we can write down a first order action
\be
{\rm e}^V\bu^i[\eta_{im}+\eta_{ik}(V_1-4\hc V)(T_1)^k{}_m]({\rm e}^{V_2T'_2})^m{}_j\uu^j
-\tilde{\eta}_1V_1-\tilde{\eta}_2V_2~,
\label{lp1}
\ee
which, after shifting $V_1$ and $V_2$, can be rewritten as
\be
{\rm e}^{\ft13V_2}(X_0-X_3+V_1X_1)+{\rm e}^{-\ft23V_2+3V}X_3-
\tilde{\eta}_1(V_1+4\hc V)-\tilde{\eta}_2(V_2-3V)~,~~ X_3\equiv \bu^3\uu^3~.
\ee
Varying with respect to $V,V_1,V_2$, we find
\begin{eqnarray}
&&3\,{\rm e}^{-\ft23V_2+3V}X_3 ~=~-3\tilde{\eta}_2+4\hc\tilde{\eta}_1~,\nonumber\\ \nonumber\\
&&e^{\ft13V_2}X_1 ~=~ \tilde{\eta}_1~,\nonumber\\ \nonumber\\
&&\frac13\left({\rm e}^{\ft13V_2}(X_0-X_3+V_1X_1)\right)-\frac23
\left({\rm e}^{-\ft23V_2+3V}X_3\right) ~=~ \tilde{\eta}_2~.
\end{eqnarray}
Again, we can solve for $V,V_1,V_2$, and again we drop numerous total
derivative terms to obtain:
\be
(\tilde{\eta}_2-4\hc\tilde{\eta}_1)\ln{(\tilde{\eta}_2-4\hc\tilde{\eta}_1)}-\tilde{\eta}_2\ln\tilde{\eta}_1~. \label{G2}
\ee

This is a deformation of another form of the universal hypermultiplet 
that was written in \cite{DWRV1}. It can be put into a slightly nicer form by
the redefinition $\tilde{\eta}_2\to\tilde{\eta}_2+4\hc\tilde{\eta}_1$:
\be
G=\tilde{\eta}_2 \ln {\frac{\tilde{\eta}_2}{\tilde{\eta}_1}} - 2 {\hat \chi}\, \tilde{\eta}_1 \ln \tilde{\eta}_1\ ,
\ee
where we have also taken into account the rescaling $\tilde{\eta}_1 \rightarrow \tilde{\eta}_1/2$ for the same reasons as in Section 3.2. Notice that the deformation term is of exactly the same form as the one in \eqn{def-G}. 
It would be interesting to understand the action of the Heisenberg algebra
in this basis, in particular the corresponding symmetries on the 
tensor multiplets $\tilde{\eta}_1$ and $\tilde{\eta}_2$.


\end{document}